\documentstyle[12pt]{article}
\newcommand{\EQ}{\begin{equation}}
\newcommand{\EN}{\end{equation}}

\renewcommand{\thefootnote}{\fnsymbol{footnote}}
\def\aprle{\buildrel < \over {_{\sim}}}
\def\aprge{\buildrel > \over {_{\sim}}}
\begin{document}
\topmargin 8mm
\oddsidemargin=-0.1truecm
\evensidemargin=-0.1truecm
\newpage
\setcounter{page}{0}
\begin{flushright}
FTUV/94-9\\
IFIC/94-6
\end{flushright}
\begin{center}
{\large ATMOSPHERIC NEUTRINOS}
\footnote{Invited talk at the Intern. School on Cosmological Dark Matter,
Valencia, Spain, October 4-8, 1993}
\end{center}
\vspace{0.2cm}
\begin{center}
{\large E.Kh. Akhmedov}
\footnote{On leave of absence from National Research Center
``Kurchatov Institute'', 123182 Moscow, Russia}\\
\normalsize
\vspace{.8cm}
{\em Instituto de Fisica Corpuscular (IFIC-CSIC) \\
Departamento de Fisica Teorica\\
Universitat de Valencia\\
c/ Dr. Moliner, 50\\
46100 Burjassot (Valencia) Spain}\\
\end{center}


\begin{abstract}
Recent results of atmospheric neutrino experiments are reviewed and their
possible interpretations are discussed, main emphasis being put on the
neutrino oscillation hypothesis.
\end{abstract}
\renewcommand{\thefootnote}{\arabic{footnote}}
\setcounter{footnote}{0}
\section{Introduction}
Atmospheric neutrinos provide us with a unique tool of studying neutrino
properties in a wide range of neutrino energies and for vastly differing
distances between the neutrino birthplace and the detector. They also
constitute a background in the underground proton decay experiments. These
are electron and muon neutrinos and their
antineutrinos which are produced in the hadronic showers induced by primary
cosmic rays in the earth's atmosphere. Their energies range from a few hundred
MeV to more then 100 TeV. The main mechanism of production of the
atmospheric neutrinos is given by the following chain of reactions:
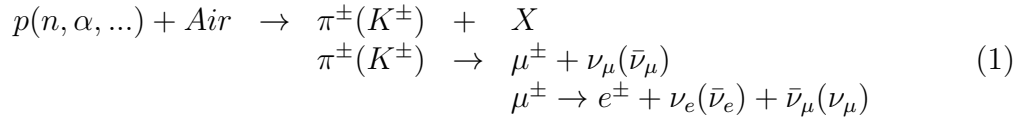
\EQ
\begin{array}{lllll}
p(n, \alpha, ...)+Air&\rightarrow &\pi^{\pm}(K^{\pm})&+ & X  \\
&   &\pi^{\pm}(K^{\pm})&\rightarrow &\mu^{\pm}+\nu_{\mu}(\bar{\nu}_{\mu})\\
& & & &\mu^{\pm}\rightarrow e^{\pm}+\nu_{e}(\bar{\nu}_{e})+
\bar{\nu}_{\mu}(\nu_{\mu})
\end{array}
\EN
Atmospheric neutrinos can be observed
directly in large mass underground detectors by means of their charged-current
interactions:
\EQ
\nu_{e}(\bar{\nu}_{e})+A\rightarrow e^{-}(e^{+})+X
\EN
\EQ
\nu_{\mu}(\bar{\nu}_{\mu})+A\rightarrow \mu^{-}(\mu^{+})+X
\EN
The atmospheric neutrino events can be subdivided into several groups,
depending on the energy of the charged leptons produced.
 ``Contained events'' are those where the neutrino--nucleus interaction vertex
is located inside the detector and all final state particles do not get out
from it. These events have charged lepton energies in the range from
a few hundred MeV to 1.2 GeV. Muon neutrinos can also be detected indirectly
by observing the muons that they have produced in the material surrounding the
detector. To reduce the background from atmospheric muons, only upward--going
neutrino-induced muons are usually considered. A rough estimate of the energy
spectrum of the upward--going muons has been obtained dividing them in two
categories, passing (or through-going) and stopping muons. The latter, which
stop inside the detector, have the energies ranging typically from 1.2 to 2.5
GeV, whereas the through-going muons have the energies $\aprge 2.5$ GeV.

Recently there has been a considerable interest in atmospheric neutrinos
due to an anomaly observed in the contained neutrino
induced events by the Kamiokande${}^{1,2}$ and IMB${}^{3}$
collaborations. Both groups measure the ratio of muons to electrons
and find it smaller than what is predicted by
several independent theoretical calculations${}^{4}$.
A possible interpretation of the effect has been given in terms of
$\nu_\mu \leftrightarrow \nu_x$ oscillations, where $\nu_x$ can be
$\nu_e$, $\nu_\tau$ or a sterile neutrino $\nu_s$, with typical values of
the oscillation parameters $\Delta m^2 \sim 10^{-3} \div 10^{-2}~{\rm eV^2}$
and $\sin^2 2\theta \ge 0.5$ (a summary is contained in ref.${}^{5}$).
For this range of parameters, the flux of upward--going muons could also be
observably reduced. However, the IMB, Baksan and Kamiokande experiments did
not observe any such reduction${}^{6,7,8}$, thus setting
rather stringent limits to the allowed region in the ($\Delta m^2$,
$\sin^2 2\theta$) plane. In this Lecture we will discuss various aspects of
the atmospheric neutrino anomaly and its possible interpretations.

\section{Predicted atmospheric neutrino fluxes and the data}
\noindent {\large {\em Contained events}}

Naively, from the reaction chain of Eq. 1 one would expect to have two
atmospheric muon neutrinos or antineutrinos for every electron neutrino or
antineutrino. In reality, this is not quite true: one should take into
account the differences in the lifetimes of $\pi^{\pm}$, $K^{\pm}$ and
$\mu^{\pm}$ as well as the differences in their spectra. Also, although the
reaction chain (1) is the dominant source of atmospheric neutrinos, it is
not the only one. Accurate calculation of the atmospheric neutrino fluxes
is a difficult job which includes such ingredients as spectra and chemical
composition of cosmic rays (including geomagnetic effects and solar
activity), cross sections of $\pi$ and $K$ production off the nuclear
targets, Monte Carlo simulation of hadronic cascades in the atmosphere and
the calculation of neutrino spectra including muon polarization effects.
Each step introduces some uncertainty in the calculation. Below we
reproduce a table of the estimated uncertainties which we borrowed from
a recent  talk of Todor Stanev${}^{9}$:\\

\underline{Contained events:} \hspace{3.5cm} {\underline
{Upward going muons:}}\\

\vspace{-0.3cm}
\begin{tabular}{lclc}
$\bullet$ cosmic ray flux             & $\sim$ 10\%
&$\bullet$ cosmic ray flux             & \\
$\bullet$ solar cycle modulation      & $\sim$ 10\%
&$\;\;$ and composition                & $\sim$ 15\% \\
$\bullet$ geomagnetic cutoff          & $\sim$ 5\%
&$\bullet$ inclusive $\pi$ spectra     & $\sim$ 12\% \\
$\bullet$ $K/\pi$ ratio               & $\sim$ 3\%
&$\bullet$ $K/\pi$ ratio               & $\sim$ 12\% \\
$\bullet$ $\nu/\bar{\nu}$ ratio       & $\sim$ 5\%
&$\bullet$ $\nu/\bar{\nu}$ ratio       & $\sim$ 8\%  \\
$\bullet$ $\nu N$ cross sections      & $\sim$ 5\%
&$\bullet$ $\nu N$ cross sections      & $\sim$ 10\%
\end{tabular}

The overall uncertainty of the calculated atmospheric neutrino fluxes is
pretty large, and the total fluxes calculated by different authors differ by
as much as 30\%. At the same time, the ratio of the muon to electron neutrino
fluxes is fairly insensitive to the above uncertainties, and all the
calculations taking into account the muon polarization effect yield the ratios
of muon-like to electron-like contained events which agree to better than 5\%.
This ratio has been measured up to now by five experimental groups. Two of
them (Kamiokande and IMB) use large water \v{C}erenkov detectors, whereas the
other three detectors (Fr\'{e}jus, NUSEX and Soudan 2) are iron calorimeters.
Muons can be distinguished from electrons in water detectors by the rings of
the \v{C}erenkov light they produce. Another way of identifying the muon-like
events is by detecting the muon decay, and the results of both identification
techniques are in a very good agreement with each other. Below we summarize
the results of the five experiments for the double ratio,
$R\equiv \frac{(\mu/e)_{data}} {(\mu/e)_{MC}}$ in the contained events ($MC$
stands for the Monte Carlo simulations):

\begin{tabular}{lccc}
Experiment       & Exposure ($kt\cdot yr$)   &  $R$\\
{}    &   {}               & {} & {} \\
Kamiokande${}^{10}$ &   6.18         & $0.60\pm 0.06\pm 0.05$ & (rings) \\
{}                      &   {}           & $0.61\pm 0.07$         & (decays)\\
IMB${}^{3}$       &  7.7           & $0.54\pm 0.05\pm 0.12$ & (rings)\\
{}                      &   {}           & $0.65\pm 0.05$         & (decays)\\
Fr\'{e}jus${}^{11}$  &  1.56          & $0.87\pm 0.21$         & {}  \\
NUSEX${}^{12}$      &  0.40          & $0.99\pm 0.40$         & {}  \\
Soudan 2 ${}^{13}$    &  1        & $0.69\pm 0.19\pm 0.09^{*)}$ & {}  \\
\end{tabular}

\small{${}^{*)}$ preliminary result}

{}From this table we see that the data from the water \v{C}erenkov detectors
show too small a $(\mu/e)$ ratio as compared to the Monte Carlo simulations,
and the results of Kamiokande and IMB groups are in a very good agreement.
This is the essence of what is called now ``the atmospheric neutrino
anomaly''.
At the same time, the iron calorimeters give the double ratio $R$ which is
consistent with one. Are these two sets of data inconsistent with each other?
The answer is no, since the calorimeter results have lower statistical
significance and in fact are consistent with the suppressed $(\mu/e)$ ratio
as well.

Now the question is, if the suppression of this ratio is a real effect, is
the muon neutrino flux suppressed or the electron neutrino flux enhanced?
The interpretation is calculation-dependent. Below we reproduce the
comparison of the Kamiokande data for double ratio $R$ with Monte Carlo
simulations using different neutrino fluxes${}^{10}$:

\noindent
Gaisser--Stanev:
$$ \frac{(\mu/e)_{data}}{(\mu/e)_{MC}}=\frac{(191/198)}{(325/203)} =
0.60\pm 0.06 \pm 0.05 $$
\noindent
Lee--Koh:
$$ \frac{(\mu/e)_{data}}{(\mu/e)_{MC}}=\frac{(191/198)}{(256/157)} =
0.59\pm 0.06 \pm 0.05 $$
\noindent
Honda {\em et al.}:
$$ \frac{(\mu/e)_{data}}{(\mu/e)_{MC}}=\frac{(191/198)}{(293/179)} =
0.59\pm 0.06 \pm 0.05 $$
\noindent
Bugaev--Naumov:
$$ \frac{(\mu/e)_{data}}{(\mu/e)_{MC}}=\frac{(191/198)}{(214/133)} =
0.60\pm 0.06 \pm 0.05 $$

We see that although the predictions for $\mu$-like and $e$-like events
obtained with different fluxes differ substantially, the calculated $(\mu/e)$
ratios are very close to each other, and so the double ratios $R$ are
practically the same. From the above numbers we can see that if one uses
the Gaisser--Stanev or Honda {\em et al.} fluxes for the interpretation of the
data, one would conclude that the observed muon neutrino flux is suppressed
whereas the electron neutrino flux is close to the calculated one. On the
contrary, the Bugaev--Naumov flux implies that the observed electron neutrino
flux is enhanced whereas that of muon neutrinos is close to the calculated
one. At the same time, the Lee--Koh flux implies that both the muon neutrino
flux is suppressed and the electron neutrino flux is enhanced. \\

\noindent {\large {\em Upward--going muons}}

Upward--going muons have been observed by several experimental groups. The
results are summarized below:

\noindent Kamiokande${}^{8}$:
$$ F_{\mu}(>3 \;\;{\rm GeV})=(2.04 \pm 0.13)\times 10^{-13}
\;\;cm^{-2}s^{-1}sr^{-1}, $$

\noindent Baksan${}^{7}$:
$$ F_{\mu}(>1 \;\;{\rm GeV})=(2.83 \pm 0.14)\times 10^{-13}
\;\;cm^{-2}s^{-1}sr^{-1}, $$

\noindent IMB${}^{6}$:
$$ F_{\mu}(>2 \;\;{\rm GeV})=(2.26 \pm 0.17)\times 10^{-13}
\;\;cm^{-2}s^{-1}sr^{-1}, $$

\noindent After the correction for different thresholds, all the results
agree well with each other.

Since there are no electrons in the upward--going flux, one has to compare
the data directly with the calculations. The results of such a comparison
for different calculated $\nu_{\mu}$ spectra and two sets of the nucleon
structure functions are presented below (muon fluxes are in the units
of $10^{-13} cm^{-2}s^{-1}sr^{-1}$ and correspond to the threshold
$E_{\mu}>3$ GeV)${}^{9}$:

\begin{tabular}{|l||c|c||c|c||c|c||c|c||} \hline
\multicolumn{1}{|l||}{Calculations} &
\multicolumn{2}{c||}{Muon Flux} &
\multicolumn{2}{c||}{$\begin{array}{c}(data/calc.)
\\ {\rm Kamiokande}\end{array}$} &
\multicolumn{2}{c||}{$\begin{array}{c}(data/calc.)\\ {\rm Baksan}
\end{array}$} &
\multicolumn{2}{c||}{$\begin{array}{c}(data/calc.)\\ {\rm IMB}\end{array}$}\\
\hline
{}  & (a) & (b) & (a) & (b) & (a) & (b) & (a) & (b) \\
{}  &  {} & {}  & {}  & {}  &  {} & {}  & {}  & {}  \\
Agraval {\em et al.}${}^{14}$ & 2.36 & 2.11 & 0.86 & 0.97 &
0.88 &0.99 & 0.81 & 0.91 \\
Butkevich {\em et al.}${}^{15}$  & 2.43 & 2.16 & 0.84 & 0.94 &
0.86 & 0.96 & 0.79 &0.89 \\
Mitsui {\em et al.}${}^{16}$ & 2.30 & 2.05 & 0.89 & 1.00 &
0.98 &1.02 & 0.83 & 094 \\
Volkova${}^{17}$       & 2.18 & 1.95 & 0.94 & 1.05 &
0.95 & 1.07 & 0.88 & 0.98 \\ \hline
\end{tabular}

Here (a) and (b) refer to the calculations with the structure functions
Owens'91 ${}^{18}$ and EHLQ2 ${}^{19}$, respectively.
Although there is a considerable spread in the calculations, one can
conclude that the data are in a good agreement with the Monte Carlo
simulations, i.e. there is no indication of the deficiency of the muon
neutrinos.

The main problem in analyzing the data on upward--going muons is the
significant uncertainty ($\sim 20\%$) in the calculations of the total fluxes.
To cure this problem, the IMB collaboration has analyzed the ratio of stopping
to passing upward--going muons. In this ratio the uncertainty related to
the unknown normalization of the flux cancels, and the result is dominated
by the statistical errors. The value of the ratio obtained was

$$\frac{F_s}{F_{p}}=0.160 \pm 0.019, $$

\noindent in an excellent agreement with their detailed Monte Carlo
calculation, $0.163 \pm 0.05$.

\section{Interpretation of the data}
{\large {\em Neutrino oscillations}}

A possible explanation of the observed anomaly in the contained atmospheric
neutrino events could be oscillations of muon neutrinos into $\nu_x$ where
$\nu_x$ can be  $\nu_e$, $\nu_\tau$ or a hypothetical sterile neutrino
$\nu_s$. In the former case one would observe both a deficiency of
atmospheric $\nu_\mu$'s and an excess of $\nu_e$'s, whereas in the latter two
cases the result would be the deficiency of the $\nu_\mu$'s. The
probability of the $\nu_\mu$ oscillations in vacuum in the two--flavor
approximation is
\EQ
P(\nu_{\mu}\rightarrow \nu_x)=\sin^2 2\theta_{\mu x} \sin^2 \left (
\frac{\Delta m^2}{4E_{\nu}}L \right ) = \sin^2 2\theta_{\mu x} \sin^2 \left (
1.27\Delta m^2({\rm eV}^2)\frac{L\;({\rm km})}{E_{\nu}\;({\rm GeV})} \right ).
\EN
Here $\theta_{\mu x}$ is the relevant mixing angle, $\Delta m^2$ is the
squared mass difference of the neutrino mass eigenstates, and $L$ is the
distance travelled by the neutrinos. The results on the contained events can
therefore be represented as an ``allowed region'' on the $\sin^2 2\theta_
{\mu x}$--$\Delta m^2$ plane (see figs. 1 and 2 below).

The important question is: Do the results on upward--going muons, in which
an unsuppressed muon neutrino flux has been observed, contradict the
neutrino--oscillations hypothesis as a possible explanation of the anomaly
in the contained events? The answer is no (or not necessarily), since the
neutrinos in these two sets of data have different characteristic energies and
also travel through different distances $L$. The neutrinos giving rise to
the upward--going muons have bigger energies then those responsible for the
contained events, therefore their oscillation lengths are larger and the
oscillations may not be fully developed. This effect is partly compensated
by the fact that they travel through larger distances before reaching the
detector. As a result, the upward--going muon data constrain the allowed
ranges of the parameters of neutrino oscillations but do not exclude this
mechanism completely.

Usually the atmospheric neutrino anomaly is analyzed in terms of the
$\nu_{\mu}\leftrightarrow \nu_{\tau}$ oscillations. The reason for this is
that the $\nu_{\mu}\leftrightarrow \nu_{e}$ oscillations are now the most
popular candidate for the solution of the solar neutrino problem (the observed
deficit of solar neutrinos). This would imply that they cannot be responsible
for the atmospheric neutrino anomaly since the ranges of $\Delta m^2$ required
in the two cases do not overlap. However, the solar neutrino deficit
can be accounted for through the $\nu_e \leftrightarrow
\nu_{\tau}$ oscillations provided the neutrino masses obey the
conditions $m_{\nu_e}\approx m_{\nu_\tau} \ll m_{\nu_\mu}$ or
$m_{\nu_\mu} \ll m_{\nu_e}\approx m_{\nu_\tau}$. It should be emphasized
that although many popular models of neutrino mass generation predict the
neutrino mass hierarchy $m_{\nu_e}  \ll m_{\nu_\mu} \ll m_{\nu_\tau}$,
models with quite different hierarchies also exist. Moreover, there are
absolutely no experimental indications in favour of the direct hierarchy.
On top of that, the solution of the solar neutrino problem may have nothing
to do with neutrino oscillations. For this reason one should consider all
possible kinds of the $\nu_{\mu}$ oscillations as possible explanations of
the atmospheric neutrino anomaly.

It is well known by now that the probability of neutrino oscillations in
matter can differ significantly from that in the vacuum${}^{20}$.
Therefore one should take into account the matter effects on the
oscillations of neutrinos passing through the earth. As we already mentioned,
the simplest possibility is that the atmospheric neutrino anomaly is due to
the $\nu_\mu \leftrightarrow \nu_\tau$  oscillations.
In this  case, since both neutrino flavors have identical interactions
with matter, the  oscillations will not be affected by the presence of
the matter in the earth, and the simple formula of Eq. 4 applies. If, however,
the initial $\nu_\mu$ oscillates to $\nu_e$ or to a sterile neutrino, $\nu_s$,
the matter effects can be relevant. The oscillation probability can then be
found by numerically solving the neutrino evolution equation.

For matter effects to be significant one needs${}^{20}$
\EQ
{\Delta m^2\over E_{\nu}} \aprle \sqrt{2} G_F N_A \rho
= 0.758 \times 10^{-4} \rho({\rm g\,cm^{-3})  ~ {eV^2 \over GeV}}
\EN
Neutrinos giving rise to the contained events have energies $E_\nu \le 1.2$
GeV, with a rapidly falling spectrum. As we shall see, the
neutrino--oscillations solution requires the absolute value of the squared
mass difference to be always larger than $10^{-3}~{\rm eV}^2$. The maximum
density of the earth is $\rho \simeq 12.5$ g\,cm$^{-3}$, therefore for
contained events matter effects are  practically negligible.

Matter effects are on the contrary significant for upward--going muons,
the reason being that the neutrino energies involved are one to two orders
of magnitude larger  than those for contained events.

The matter effects on the oscillations of neutrinos responsible for the
upward--going muons have been considered in${}^{21}$.  The results for
the $\nu_{\mu}\leftrightarrow \nu_{\tau}$ and $\nu_{\mu}\leftrightarrow
\nu_{s}$ oscillations, along with the constraints from the other
experiments, are summarized in fig. 1.
In this figure three limiting curves are the results of experiments that do
{\it not} involve upward--going muons.
Curve (a)  is obtained from accelerator experiments${}^{22}$ and the
allowed region is below the curve. Curve (b) is obtained from the result of
Fr\'ejus${}^{11}$ on the $e/\mu$ ratio of contained events; this
result is consistent with the no--oscillation hypothesis, and the allowed
region is to the left of the curve. Curve (c) is obtained from the observation
of  an anomaly in the same $e/\mu$ ratio by Kamiokande and IMB${}^{1,3}$;
this is a ``positive result'' and the allowed region is
to the right of the curve. These three limits apply equally to the
$\nu_{\mu}\leftrightarrow \nu_{\tau}$ and $\nu_{\mu}\leftrightarrow
\nu_{s}$ oscillations which differ from each other only because of the matter
effects. These effects are obviously irrelevant to the accelerator limit, and
almost so also for (b) and (c) curves, although approximately half of the
contained events observed in IMB and Kamiokande are produced by neutrinos that
have penetrated through the earth. This fact has already been explained
above (see the discussion after Eq. 5).

The limits obtained  from the total flux of upward--going muons using the
criterion${}^{21}$
\EQ
F\leq 0.75 \;F_0,
\EN
$F_{0}$ being the average flux calculated in the absence of the oscillations,
are shown in fig. 1 by the  curves $A_\tau$, $A_{s^+}$ and $A_{s^-}$,
the subscript indicating the flavour of the neutrino mixed with the $\nu_\tau$.
The case $\nu_s$ is described by two curves because one needs to consider the
sign of $\Delta m^2$ whenever the matter effects on neutrino oscillations
are important.

The limits obtained  from the stopping/passing ratio
using the criterion${}^{21}$
\EQ
\frac{F_s}{F_p}\leq 0.8\;\left (\frac{F_s}{F_p}\right )_0,
\EN
where the subscript ``0'' stands for the ratio of fluxes in the absence of the
oscillations, are shown in the same figure by the curves $B_\tau$, $B_{s^+}$
and $B_{s^-}$.

Comparing the three type $A$ curves we can make the following observations.
When $\Delta m^2 < 0$
($\Delta m^2 > 0$) for oscillations of neutrinos (antineutrinos) the MSW
resonant enhancement${}^{20}$ will occur. Since $\nu_\mu$'s are more
abundant (and have a larger cross section) than $\bar{\nu}_\mu$'s, the same
values of $\sin^2 2 \theta$ and  $|\Delta m^2|$ will result in a stronger
suppression of the upward--going muon flux if $\Delta m^2$ is negative and
the MSW  resonance is relevant for neutrinos. For $|\Delta m^2|
\aprge 1~{\rm eV^2}$ the three curves $A_\tau$, $A_{s^+}$ and $A_{s^-}$ almost
coincide; this is a reflection of the fact that for large
$|\Delta m^2|/E$ matter effects are not significant.

Very similar considerations can be made about the curves $B_\tau$, $B_{s^+}$
and $B_{s^-}$. For maximal mixing, the region of $|\Delta m^2|$ that can be
excluded in the case of $\nu_{\mu}\leftrightarrow \nu_{s}$ oscillations
is moved to values of $|\Delta m^2|$ larger by a factor 2.8  than in the
$\nu_{\mu}\leftrightarrow \nu_{\tau}$ case.  For oscillations into sterile
neutrinos, when the mixing is smaller than unity, the limit that applies
for $\Delta m^2 < 0$ is stronger than  for the other case. This is again
because when  the  MSW resonance occurs for neutrinos (antineutrinos) the
sensitivity to the oscillations is stronger (weaker). A detailed analysis
must also take into account the fact that when the neutrinos go through the
oscillation resonance the oscillation length passes through a maximum. This is
the reason why the limit $B_{s^+}$ is stronger than the limit $B_{s^-}$ in a
limited  region of parameter space. For maximal mixing the curves $B_{s^+}$
and $B_{s^-}$ end in the same points.

The most significant difference between the limits for the
$\nu_{\mu}\leftrightarrow \nu_{\tau}$ and $\nu_{\mu}\leftrightarrow \nu_{s}$
oscillations is in the region of low squared mass differences. A region of
the parameter space $|\Delta m^2| = (3 \div 7) \times 10^{-4} ~{\rm eV^2}$
and $\sin^2 2 \theta \ge 0.7$  is excluded for the
$\nu_\tau$ but not for $\nu_s$ oscillations. This region is however
already excluded by curve (c), provided that the anomaly in the $e/\mu$ ratio
of contained events is due to neutrino oscillations. As shown in fig. 1,
for both types of oscillations there is a region of parameter space
which is  compatible with all existing experimental measurements. The
range of allowed values  for the parameters is  in both cases roughly:
$0.4 \aprle \sin^2 2 \theta \aprle 0.7$ and $2 \times 10^{-3}\;({\rm eV^2})
\aprle |\Delta m^2| \aprle 0.4\;({\rm eV^2})$.
The allowed region  for $\nu_s$ is somewhat smaller but contains also a small
part (for large positive $\Delta m^2$) which is not allowed for $\nu_\tau$.
In future, with more data, the sensitivity of the curves of type $B$ (which is
determined by the statistical errors) will improve, exploring the low
$|\Delta m^2|$ part of the allowed region. To improve the sensitivity of the
curves of type $A$, one needs rather to reduce the theoretical systematic
uncertainties. If these could be reduced to the level of 15\% (at 90\% c.l.)
the entire allowed region of parameter space would then be explored by
measurements of upward--going muons and one would be able to either confirm
or disprove the neutrino oscillations as a solution to the atmospheric
neutrino puzzle.

As we have seen, the oscillations of $\nu_\mu$ into a sterile neutrino state
$\nu_s$ can be strongly affected by the matter of the earth. One should
notice that although the existence of a sterile neutrino does not contradict
any laboratory data, it can be in
conflict with cosmological considerations. Namely, for large enough
values of the mixing angle and $\Delta m^2$, $\nu_{\mu}\leftrightarrow
\nu_s$ oscillations can bring the sterile neutrinos into equilibrium with
matter before the nucleosynthesis epoch thereby affecting the primordial
${}^4$He abundance in the universe.  The analysis sets the maximum
allowed number of ``light neutrinos'' $N_\nu$ to 3.4. In order not have
$N_\nu > 3.4$  for oscillations between sterile and muon neutrinos
one needs $\Delta m^2 < 8\times 10^{-6}\;{\rm eV}^2$ for maximal mixing
(see${}^{23}$ for more details).
Clearly, with the values $|\Delta m^2 |>2\times 10^{-3}\;{\rm eV}^2$ this
bound would be violated, and one would have $N_\nu = 4$.

The situation about the experimentally allowed values of the oscillation
parameters in the case of the $\nu_{\mu}\leftrightarrow \nu_{e}$ oscillations
is summarized in fig. 2. The results from the analysis of solar neutrino
experiments refer to lower values of $|\Delta m^2|$ and are not shown.
As in fig. 1, three limiting curves are the results of experiments that do
{\it not} involve upward--going muons.
Curve (a)  is the limit obtained from reactor experiments${}^{24}$,
curve (b) and (c) are  obtained from  measurements of the
$e/\mu$ ratio of  contained events
in the Fr\'ejus${}^{11}$ and Kamiokande and IMB experiments${}^{1,3}$.
As before, curve (b) excludes the region to its  right, curve (c) the
region to its left.

The limits obtained  from the measurement of the total upward--going
muon flux according to the criterion of Eq. 6 is shown by
curves $A_{e^+}$ and $A_{e^-}$, the two curves referring to the
sign of $\Delta m^2$ as before. The region excluded by the curves of type $A$
is well inside the region of parameter space already excluded by the G\"osgen
reactor experiment. Matter suppresses the oscillations more strongly
than in the $\nu_{\mu}\leftrightarrow \nu_{s}$ case, since the difference in
the effective potentials is twice as large now${}^{21}$. One also has to
take into  account the fact that in cosmic ray showers a $\nu_e(\bar{\nu}_e)$
flux is produced as well. Both effects reduce the sensitivity of the
measurement to the $\nu_{\mu}\leftrightarrow \nu_{e}$ oscillations.

For the $\nu_{\mu}\leftrightarrow \nu_{e}$ oscillations, the condition of
Eq. 7 is never satisfied${}^{21}$ and therefore no limit can be obtained
from  the stopping/passing ratio. In the case of maximal mixing, $F_s/F_p$
reaches (for $|\Delta m^2| = 6 \times 10^{-2} \;{\rm eV^2}$) a minimum value
of 0.150, only 19\% smaller than the value calculated in the absence of
oscillations. Curves $B_{e^+}$ and $B_{e^-}$  are calculated with the more
demanding criterion $F_s/F_p  \le  0.9~ (F_s/F_p)_0$. They are an indication
of the sensitivity that could be obtained  with a sample of data approximately
four times the one collected by IMB.

As shown in fig. 2, a small region of parameter space $0.35 \aprle \sin^2 2
\theta \aprle 0.7$ and $4\times 10^{-3}\;({\rm eV^2}) \aprle |\Delta m^2|
\aprle 2 \times 10^{-2}\;({\rm eV^2})$ is compatible with all existing
experimental measurements.

As we have seen the complex of data induced by atmospheric neutrinos
(contained and upward--going muons) may be described in terms of neutrino
oscillations. The matter effects are important in the precise determination
of the allowed parameter region. \\

\noindent {\large {\em Proton decay}}

Another possible explanation of the anomaly observed in the contained
events in underground detectors is the decay${}^{25}$
\EQ
p\rightarrow e^{+}+\nu_{e}+\bar{\nu}_{e}.
\EN
Such a decay would produce an excess of electron neutrinos. Obviously,
the decay neutrinos would have the energy $<1$ GeV and so the data on
the upward--going muons would not be affected. The contained data can be
accounted for provided the proton lifetime with respect to the process (8)
is $\tau_p=\left(3.7^{+1.9}_{-1.0}\right )\times 10^{31}$ yr. One should
keep in mind that although the existing lower limits on the proton lifetime
seem to exclude such short a lifetime, these limits actually do not apply to
the exotic decay mode of Eq. 8, and so the proton decay is still a viable
explanation of the atmospheric neutrino anomaly. This possibility can be
checked experimentally. The Soudan 2 detector is
capable of detecting the proton recoil in the process
\EQ
\nu_{e}+N\rightarrow e^{-}+p.
\EN
If the proton decay hypothesis is correct, a fraction of the
electron-like events should be due to the decay (8), and not to the
reaction (9); thus one should see fewer recoil protons then expected in the
electron-like events${}^{13}$. So far, Soudan 2 have not seen any
deficiency of recoil protons in their data. New data with higher statistics
will probably shed some light on this exciting possibility.

\section{Summary and outlook}
The contained event data in the underground water \v{C}erenkov detectors
shows a reduced $(\mu/e)$ ratio implying a deficiency of atmospheric
$\nu_{\mu}$'s or an excess of $\nu_{\mu}$'s. The results of iron calorimeters
are consistent with both the reduced and unsuppressed $(\mu/e)$ ratio
because of their statistics being too low.

Possible explanations of the observed anomaly include neutrino oscillation
and proton decay. All kinds of neutrino oscillations ($\nu_{\mu}
\leftrightarrow \nu_{\tau}$, $\nu_{\mu}\leftrightarrow \nu_{e}$, $\nu_{\mu}
\leftrightarrow \nu_{s}$) can account for the atmospheric neutrino anomaly.
The data on upward going muons severely constrain the allowed  $(\Delta m^2,
\sin^2 2\theta)$ space for neutrino oscillations but do not exclude them
as a possible solution of the problem. Matter effects on neutrino oscillations
are important for the precise determination of the allowed parameter region.

There are a number of theoretical works yielding the neutrino masses and
mixings which can simultaneously explain the atmospheric neutrino
anomaly and the solar neutrino problem through the (different kinds of the)
neutrino oscillations. Upward--going muons data rules out large mixing angles
($\sin^2 2\theta >$0.6--0.8) in atmospheric neutrino oscillations, thereby
severely constraining the theoretical models.

Is the atmospheric neutrino anomaly a manifestation of new physics or an
instrumental effect? At present, we don't have a definite answer.
Still there are possibilities to improve our understanding of the problem.
For the analyses of the data accurate calculations of the atmospheric neutrino
fluxes are required.
At present the uncertainties in these calculations are rather large. To
improve the situation, new measurements of primary cosmic ray spectra and
composition would be highly desirable. The neutrino--nucleus cross sections
introduce large uncertainties in the calculated spectra ($\sim 10\%$). More
accurate cross sections are needed. For the interpretation of the data of
the \v{C}erenkov detectors, the $\mu/e$ identification probability is
crucial. Although the results obtained with two different identification
techniques agree very well with each other, direct check of the $\mu/e$
identification probability in water would be welcome. The Kamiokande
collaboration is planning to perform the direct experiment using the KEK
Proton-Synchrotron with $\mu$, $e$ and $\pi$ beams with the momenta
$p=$200--1000 MeV/$c$. The 1000 $t$ detector is ready, and the experiment was
scheduled to start (and probably started) in December 93.

To check the neutrino oscillations hypothesis, direct long--baseline
accelerator experiments covering the interesting oscillation parameter range
would be of crucial importance. There were several proposals of such
experiments, including the Super--Kamiokande and Icarus experiments using
the neutrino beams from CERN and the Soudan 2 experiment using the FNAL
beam. Recently another experiment, using high intensity neutrino beam of the
AGS accelerator of the Brookhaven National Laboratory, has been proposed.

Thus, much work has to be done to clear up the situation with the atmospheric
neutrino anomaly.

It is a pleasure to thank Paolo Lipari and Maurizio Lusignoli in
collaboration with whom the work${}^{21}$ has been done. Useful discussions
with Zurab Berezhiani and Alexei Smirnov are gratefully acknowledged.


\newpage
\noindent {\bf Figure Captions}
\begin{itemize}
\item [Fig. 1]
Limits for the oscillation parameters
$\Delta m^2$ and $\sin^2 2 \theta$ in the case of
$\nu_\mu \leftrightarrow \nu_\tau$ and
$\nu_\mu \leftrightarrow \nu_s$   mixing.
The curves $A_\tau$, $A_{s^+}$ and $A_{s^-}$ are
limits  obtained from the  measurements of the  total flux
of upward--going muons.
The curves $B_\tau$, $B_{s^+}$ and $B_{s^-}$ are
obtained from the measurements of the stopping/passing ratio.
Also plotted are the 90\% c.l. limits from accelerator experiments
(a),  and  the from the measurement of the
$e/\mu$ ratio of contained events in  the Fr\'ejus (b) and
Kamiokande experiments (c).
\item [Fig. 2.]
Limits for the oscillation parameters
$\Delta m^2$ and $\sin^2 2 \theta$ in the case of
$\nu_\mu \leftrightarrow \nu_e$   mixing.
The curves $A_{e^+}$ and $A_{e^-}$ are
obtained from the  measurements of the  total flux
of upward--going muons.
The curves $B_{e^+}$ and $B_{e^-}$ are in this case
estimates of the sensitivity of future measurements
of the stopping/passing ratio (see text).
Also plotted are the 90\% c.l.
limits from the G\"osgen reactor experiment (a), and from the
$e/\mu$ ratio of contained events in  Fr\'ejus (b),
and  Kamiokande (c).
\end{itemize}
\end{document}